\makeatletter \def\input@path{{styles/}{styles/}} \makeatother

\ifx\JoCGVersion\undefined
    \documentclass[12pt]{article}%
    \def\UseBibLatex{1}%
    \providecommand{\JoCG}[1]{}%
    \providecommand{\NotJoCG}[1]{#1}%
\else
    \documentclass{jocg}
    \providecommand{\JoCG}[1]{#1}%
    \providecommand{\NotJoCG}[1]{}%
    \usepackage{lineno}
    \linenumbers%
\fi

\usepackage{caption}

\ifx\sarielComp\undefined%
\newcommand{\SarielComp}[1]{}
\newcommand{\NotSarielComp}[1]{#1}%
\else
\newcommand{\SarielComp}[1]{#1}%
\newcommand{\NotSarielComp}[1]{}%
\fi

\newcommand{\UsePackage}[1]{%
  \IfFileExists{styles/#1.sty}{%
      \usepackage{styles/#1}%
   }{%
      \IfFileExists{../styles/#1.sty}{%
         \usepackage{../styles/#1}%
      }{%
         \usepackage{#1}%
      }%
   }%
}

\usepackage{graphicx}%
\NotJoCG{%
   \usepackage[in]{fullpage}%
}

\usepackage{amsmath}%
\usepackage{amsfonts}%
\usepackage{amssymb}%
\usepackage{xcolor}%
\usepackage{caption}%

\usepackage{euscript}%

\SarielComp{\usepackage{sariel_colors}}%

\NotJoCG{%
   \usepackage[amsmath,thmmarks]{ntheorem}%
   \theoremseparator{.}%

   \usepackage{titlesec}%
   \titlelabel{\thetitle. }%
}
\usepackage{xcolor}%
\usepackage{mleftright}%
\usepackage{xspace}%
\usepackage{hyperref}%

\UsePackage{mathcalb}

\usepackage{hyperref}%
\hypersetup{%
      unicode,
      breaklinks,%
      colorlinks=true,%
      urlcolor=[rgb]{0.25,0.0,0.0},%
      linkcolor=[rgb]{0.5,0.0,0.0},%
      citecolor=[rgb]{0,0.2,0.445},%
      filecolor=[rgb]{0,0,0.4},
      anchorcolor=[rgb]={0.0,0.1,0.2}%
}
\usepackage[ocgcolorlinks]{ocgx2}

\NotJoCG{%
\theoremseparator{.}%

\theoremstyle{plain}%
\newtheorem{theorem}{Theorem}[section]

\newtheorem{lemma}[theorem]{Lemma}

\newtheorem{fact}[theorem]{Fact}

\newtheorem{invariant}[theorem]{Invariant}
\newtheorem{question}[theorem]{Question}

\newtheorem{prop}[theorem]{Proposition}
\newtheorem{openproblem}[theorem]{Open Problem}

\theoremstyle{plain}%
\theoremheaderfont{\sf} \theorembodyfont{\upshape}%
\newtheorem*{remark:unnumbered}[theorem]{Remark}%
\newtheorem{remark}[theorem]{Remark}%

\newtheorem{defn}[theorem]{Definition}

\newtheorem{problem}[theorem]{Problem}

\newtheorem{exercise_h}[theorem]{Exercise}
\newtheorem{assumption}[theorem]{Assumption}%

\newcommand{\myqedsymbol}{\rule{2mm}{2mm}}

\theoremheaderfont{\em}%
\theorembodyfont{\upshape}%
\theoremstyle{nonumberplain}%
\theoremseparator{}%
\theoremsymbol{\myqedsymbol}%
\newtheorem{proof}{Proof:}%

\newtheorem{proofof}{Proof of\!}%
}

\JoCG{%
   \usepackage{amsthm}

   \theoremstyle{plain}
   \newtheorem{theorem}{Theorem}[section]
   \newtheorem{lemma}[theorem]{Lemma}

   \theoremstyle{remark}%

   \newtheorem{remark}[theorem]{Remark}%

   \theoremstyle{definition}
   \newtheorem{defn}[theorem]{Definition}

}%

\definecolor{blue25emph}{rgb}{0, 0, 11}
\providecommand{\emphic}[2]{%
   \textcolor{blue25emph}{%
      \textbf{\emph{#1}}}%
   \index{#2}}
\JoCG{%
\renewcommand{\emphic}[2]{%
   {%
      \textbf{\emph{#1}}}%
   \index{#2}}
}
\providecommand{\emphi}[1]{\emphic{#1}{#1}}

\definecolor{almostblack}{rgb}{0, 0, 0.3}
\providecommand{\emphw}[1]{{\textcolor{almostblack}{\emph{#1}}}}%
\JoCG{\renewcommand{\emphw}[1]{\emph{#1}}}%

\newcommand{\atgen}{\symbol{'100}}
\newcommand{\SarielThanks}[1]{\thanks{Department of Computer Science;
      University of Illinois; 201 N. Goodwin Avenue; Urbana, IL,
      61801, USA; {\tt sariel\atgen{}illinois.edu}; {\tt
         \url{http://sarielhp.org/}.} #1}}

\newcommand{\GemmaThanks}[1]%
{%
   \thanks{%
      Department of Computer Science and Engineering; Hong Kong
      University of Science and Technology; Clear Water Bay, Kowloon,
      Hong Kong; {\tt zgaoao\atgen{}connect.ust.hk}. #1 }
}

\newcommand{\HLink}[2]{\hyperref[#2]{#1~\ref*{#2}}}
\newcommand{\HLinkSuffix}[3]{\hyperref[#2]{#1\ref*{#2}{#3}}}

\newcommand{\figlab}[1]{\label{fig:#1}}
\newcommand{\figref}[1]{\HLink{Figure}{fig:#1}}

\newcommand{\thmlab}[1]{{\label{theo:#1}}}
\newcommand{\thmref}[1]{\HLink{Theorem}{theo:#1}}

\providecommand{\deflab}[1]{\label{def:#1}}
\newcommand{\defref}[1]{\HLink{Definition}{def:#1}}
\newcommand{\defrefY}[2]{\hyperref[def:#2]{#1}}

\newcommand{\seclab}[1]{\label{sec:#1}}
\newcommand{\secref}[1]{\HLink{Section}{sec:#1}}

\newcommand{\lemlab}[1]{\label{lemma:#1}}
\newcommand{\lemref}[1]{\HLink{Lemma}{lemma:#1}}%

\providecommand{\eqlab}[1]{}%
\renewcommand{\eqlab}[1]{\label{equation:#1}}
\newcommand{\Eqref}[1]{\HLinkSuffix{Eq.~(}{equation:#1}{)}}

\newcommand{\remove}[1]{}%
\newcommand{\Set}[2]{\left\{ #1 \;\middle\vert\; #2 \right\}}
\newcommand{\pth}[2][\!]{\mleft({#2}\mright)}%

\newcommand{\ceil}[1]{\left\lceil {#1} \right\rceil}
\newcommand{\floor}[1]{\left\lfloor {#1} \right\rfloor}

\newcommand{\G}{\Mh{G}}%
\newcommand{\brc}[1]{\left\{ {#1} \right\}}
\newcommand{\cardin}[1]{\left| {#1} \right|}%

\renewcommand{\th}{th\xspace}

\renewcommand{\Re}{\mathbb{R}}%

\usepackage[inline]{enumitem}

\newlist{compactenumA}{enumerate}{5}%
\setlist[compactenumA]{topsep=0pt,itemsep=-1ex,partopsep=1ex,parsep=1ex,%
   label=(\Alph*)}%

\newlist{compactenuma}{enumerate}{5}%
\setlist[compactenuma]{topsep=0pt,itemsep=-1ex,partopsep=1ex,parsep=1ex,%
   label=(\alph*)}%

\newlist{compactenumI}{enumerate}{5}%
\setlist[compactenumI]{topsep=0pt,itemsep=-1ex,partopsep=1ex,parsep=1ex,%
   label=(\Roman*)}%

\newlist{compactenumi}{enumerate}{5}%
\setlist[compactenumi]{topsep=0pt,itemsep=-1ex,partopsep=1ex,parsep=1ex,%
   label=(\roman*)}%

\newlist{compactitem}{itemize}{5}%
\setlist[compactitem]{topsep=0pt,itemsep=-1ex,partopsep=1ex,parsep=1ex,%
   label=\ensuremath{\bullet}}%

\newcommand{\BibLatexMode}[1]{#1}
\newcommand{\BibTexMode}[1]{}

\usepackage[bibencoding=utf8,style=alphabetic,backend=biber]{biblatex}%
\UsePackage{sariel_biblatex}%

\numberwithin{figure}{section}%
\numberwithin{table}{section}%
\numberwithin{equation}{section}%

\newcommand{\Cube}{\Mh{\mathcal{C}}}%
\newcommand{\Cells}{\Mh{\EuScript{C}}}%
\newcommand{\diamX}[1]{\mathrm{diam}\pth{ #1}}
\newcommand{\sdX}[1]{\mathrm{sidelen}\pth{ #1}}

\newcommand{\etal}{\textit{et~al.}\xspace}
\newcommand{\cenX}[1]{\Mh{\mathsf{c}}\pth{#1}}

\newcommand{\qt}{\Mh{\mathcal{T}}}%

\newcommand{\ball}{\Mh{\mathcalb{b}}}
\newcommand{\hippo}{\Mh{\mathcalb{h}}}

\newcommand{\ballX}[1]{\ball\pth{#1}}
\newcommand{\ballY}[2]{\ball\pth{#1, #2}}

\newcommand{\hippoY}[2]{\hippo\pth{#1, #2}}

\newcommand{\Sphere}{\Mh{\mathbb{S}}}%

\newcommand{\orderset}{\Pi}
\newcommand{\ordAll}{\orderset^+}%

\newcommand{\Eps}{\Mh{\mathcal{E}}}%
\newcommand{\LgEps}{\Mh{\lambda}}

\newcommand{\DSet}{\Mh{\mathcal{D}}}%

\newcommand{\projY}[2]{\smash{\downarrow_#1}\pth{#2}}

\newcommand{\order}{\sigma}
\newcommand{\Of}{\mathcal{O}}
\newcommand{\pC}{\Mh{u}}%
\newcommand{\qte}{\mathcal{T}_\eps}

\newcommand{\permut}[1]{\left\langle {#1} \right\rangle}
\newcommand{\DotProdY}[2]{\permut{{#1},{#2}}}

\newcommand{\normX}[1]{\left\| {#1} \right\|}
\newcommand{\dY}[2]{\normX{#1 - #2}}

\newcommand{\dsY}[2]{\Mh{\mathsf{d}}\pth{#1, #2}}%
\newcommand{\dGY}[2]{\Mh{\mathcalb{g}\pth{#1, #2}}}%
\newcommand{\dsGY}[2]{\Mh{\mathcalb{g}\pth{#1, #2}}}%

\newcommand{\cc}{\Mh{\mathsf{c}}}

\newcommand{\Term}[1]{\textsf{#1}}
\newcommand{\LCA}{\Term{LCA}\xspace}%
\newcommand{\LSO}{\Term{LSO}\xspace}%
\newcommand{\LSOs}{\Term{LSO}s\xspace}%

\usepackage{stmaryrd}%
\providecommand{\IntRange}[1]{\mleft\llbracket #1 \mright\rrbracket}
\newcommand{\IRX}[1]{\IntRange{#1}}%

\providecommand{\TPDF}[2]{\texorpdfstring{#1}{#2}}

\newcommand{\remlab}[1]{\label{rem:#1}}
\newcommand{\remref}[1]{\HLink{Remark}{rem:#1}}%

\newcommand{\GL}{\Mh{\mathcal{L}}}

\providecommand{\Mh}[1]{#1}

\newcommand{\Grid}{\Mh{\mathcal{G}}}%
\newcommand{\Cell}{\Mh{\square}}%

\newcommand{\p}{\Mh{p}}
\newcommand{\q}{\Mh{q}}
\renewcommand{\P}{\Mh{P}}
\newcommand{\N}{\Mh{N}}
\newcommand{\Packing}{\Mh{N}}

\newcommand{\rad}{\Mh{r}}

\newcommand{\Line}{\Mh{\mathcalb{l}}}
\newcommand{\LineA}{\Mh{\mathcalb{g}}}

\newcommand{\shift}{\Mh{\nu}}

\newcommand{\OYC}{\Mh{\mathfrak{O}}}%
\newcommand{\OY}[2]{\OYC\pth{#1, #2}}%
\newcommand{\OZ}[3]{\OYC\pth{#1, #2, #3}}%
\newcommand{\seg}{\Mh{\mathsf{s}}}%

\newcommand{\precX}[1]{\prec_{#1}}
\newcommand{\preceqX}[1]{\preccurlyeq_{#1}}

\newcommand{\Forest}{\Mh{\mathcal{F}}}%
\newcommand{\Shifts}{\Mh{\mathcal{S}}}%

\newcommand{\dZ}[3]{\mathsf{d}_{#1}\pth{#2,#3}}

\newcommand{\num}{\Mh{\zeta}}%
\newcommand{\numA}{\Mh{L}}

\newcommand{\UC}{[0,1)^d}

\newcommand{\HC}{\Mh{\mathcal{H}}}%

\newcommand{\eps}{\varepsilon}%
\newcommand{\epsA}{\Mh{\xi}}%

\JoCG{%
   \newcommand{\tcite}[1]{\cite{#1}}
}

   \newcommand{\myparagraph}[1]{\paragraph*{#1.}}

   \bibliography{less_orderings}%

\title{%
       Near-Optimal Euclidean Locality-Sensitive Orderings%
   \thanks{A preliminary version of this paper has appeared in SoCG
      2024 \cite{gh-nolso-24}. The full version of the paper appeared
      in JoCG \cite{gh-noels-25}.}%
}

\author{%
   Zhimeng Gao%
   \GemmaThanks{}%
   \and%
   Sariel Har-Peled%
   \SarielThanks{Work on this paper was partially supported by NSF AF
      award CCF-2317241.  } }%

   \date{\today}%
\begin{document}

\maketitle

\begin{abstract}
    For a parameter $\eps \in (0,1)$, a set of orderings is
    $\eps$-locality-sensitive (\LSO{}s) if for any two points,
    $\p, \q \in \UC$, there exist an order in the set such that all
    the points between $\p$ and $\q$ (in the order) are $\eps$-close
    to either $\p$ or $\q$.  Since the original construction of
    \LSO{}s can not be (significantly) improved, we present a
    construction of modified \LSOs, that yields a smaller set, while
    preserving their usefulness.  Specifically, the resulting set of
    \LSOs has size $O(\Eps^{d-1} \log \Eps)$, where $\Eps =
    1/\eps$. This improves over previous work by a factor of $\Eps$,
    and is optimal up to a factor of $\log \Eps$.

    This results in a flotilla of improved dynamic geometric
    algorithms, such as maintaining bichromatic closest pair, and
    spanners, among others. In particular, for geometric dynamic
    spanners the new result matches (up to the aforementioned
    $\log \Eps$ factor) the lower bound. Thus providing a near-optimal
    simple dynamic data-structure for maintaining spanners under
    insertions and deletions.
\end{abstract}

\newcommand\VR{\rule[-0.4\baselineskip]{0.4pt}{1.2\baselineskip} }

\section{Introduction}
Given a total linear order $\order$ of $\HC = \UC$, and any two
points $\p,\q \in \HC$, for $\p \precX{\order} \q$, consider the
interval of all points between $\p$ and $\q$ according to $\order$:
\begin{equation}
    \order(\p,\q)
    =%
    \Set{ u \in \HC}{ \p \precX{\order} u \precX{\order} \q }.
    \eqlab{interval}
\end{equation}
Chan \etal \cite{chj-lota-20} showed that there is a ``small'' set
$\Pi$ of orderings, such that for every pair of points
$\p,\q \in \HC$, there is an ordering $\order \in \Pi$ that is
$\eps$-local for them, where $\eps \in (0,1)$ is a fixed parameter.

\NotJoCG{\medskip}%

\medskip%
\noindent%
\begin{minipage}{0.54\linewidth}
\begin{defn}
    \deflab{local}%
    For a pair of points $\p,\q \in \HC = \UC$, an order $\order$ over
    the points of $\HC$ is \emphi{$\eps$-local}, for $\p$ and $\q$, if
    \begin{equation*}
        \order(\p,\q)
        \quad\subseteq\quad
        \ballY{\p}{\eps \ell}
        \cup
        \ballY{\q}{\eps \ell} ,
    \end{equation*}
    where $\ell = \dY{\p}{q}$, and $\ballY{\p}{r}$ denotes the
    \emphi{ball} of radius $r$ centered at $\p$, see \figref{direct}
    and \figref{locality} (i).
\end{defn}
\end{minipage}%
\hfill\vline\hfill%
\begin{minipage}{0.39\linewidth}
    \begin{minipage}[t]{0.38\linewidth}
        \centerline{{\includegraphics[page=1,width=0.99\linewidth]{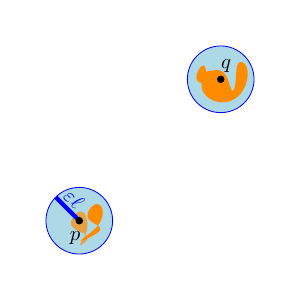}}}
    \end{minipage}
    \hspace*{-0.15\linewidth}%
    \begin{minipage}{0.7\linewidth}
        \JoCG{\bigskip}%

        \JoCG{\bigskip}%

        \JoCG{\bigskip}%

        \captionof{figure}{All the points between $\p$ and $\q$ in
           $\sigma$ (i.e., the orange colored region) are inside the
           balls around $p$ and $q$.}  \figlab{direct}
    \end{minipage}
\end{minipage}%

\NotJoCG{\bigskip}%

\noindent
Namely, all the points between $\p$ and $\q$ in $\order$ are in the
vicinities of $\p$ and $\q$ in $\HC$. Informally, there is an order in
$\Pi$ that respects the localities of $\p$ and $\q$. Chan \etal,
inspired by locality sensitive hashing, referred to $\order$ as being
a \emphw{locality sensitive ordering} (\LSO).

Chan \etal showed that one can compute such a ``universal'' set of
orderings $\Pi$, of size $O(\Eps^d \log \Eps)$, where $\Eps =
1/\eps$. This set of orderings can be easily computed, and one can
compute quickly the order of two points according to a specific order
in the set.  Having such a set of orderings reduces some dynamic
geometric problems to the maintenance of a ``few'' sorted lists of
points under insertions and deletions.  Thus, certain problems in $d$
dimensions, are reduced to a collection of problems in one
dimension. This can be interpreted as a somewhat strange dimension
reduction result, (multi)embedding $\UC$ in the real line, such that
for any two points in the hypercube, there is at least one embedding
that has certain desired geometric property.

For example, using these orderings one can easily maintain a spanner
under insertion/deletion operations with logarithmic time per
operation (ignoring terms depending on $\eps$). Specifically, one
maintains (under updates) sorted lists of the point set, according to
each of the orderings in $\Pi$. Here, an edge in the spanner is
present only if the two points are adjacent in one of these ordered
lists. Note that for each of these one dimensional instances, one
maintains a dynamic ordered map of the points (e.g., balanced binary
search tree).

\myparagraph{The challenge}

It is natural to try and reduce the number of orderings further,
improving (in hopefully a black-box fashion) the results that use
locality-sensitive orderings. Unfortunately, it is not hard (see
\lemref{old:tight}) to show that $\Omega(\Eps^d)$ is a lower bound on
the number of orderings under the $\eps$-local condition, implying
that the construction of Chan \etal is essentially optimal (ignoring
the annoying $\log \Eps$ factor).

So to reduce the number of orderings further, we have to relax the
requirement somehow. A natural first step is to enlarge the vicinity
to $\p$ and $\q$ to be an ``$\eps$-hippodrome'' region.

\begin{defn}
    \deflab{l:hippo}%
    For a segment $pq$ in $\Re^d$, and a number $r \geq 0$, let
    $\hippoY{\p \q}{r}$ be the \emphi{hippodrome} of $pq$ of radius
    $r$. Formally, it is the set of all the points in distance at most
    $r$ from $pq$:
    \begin{math}
        \hippoY{pq}{r} = \Set{u \in \Re^d}{\dsY{u}{pq} \leq r},
    \end{math}
    where $\dsY{u}{pq} = \min_{v \in pq} \dY{u}{v}$.

    For a pair of points $\p,\q \in \HC = \UC$, an order $\order$ over
    the points of $\HC$ is \emphi{$\eps$-hippodrome local}, for $\p$
    and $\q$, if
    \begin{equation*}
        \order(\p,\q)
        \quad\subseteq\quad
        \hippoY{\p\q}{\eps \ell},
    \end{equation*}
    where $\ell = \dY{\p}{q}$. See \figref{locality} (i) and (ii).
\end{defn}

\begin{figure}[t]
    \begin{tabular}{c|c|c}
      \includegraphics[page=1,width=0.3\linewidth]{figs/gap}
      &%
        \phantom{}%
        \hfill
        \includegraphics[page=2,width=0.3\linewidth]{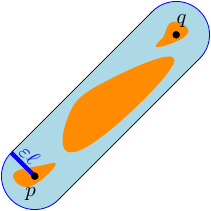} \hfill%
        \hfill
        \phantom{}%
      &
        \includegraphics[page=3,width=0.3\linewidth]{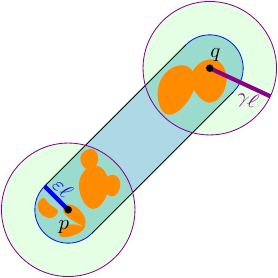}
      \\
      (i)&(ii)&(iii)
    \end{tabular}%

    \caption{The three types of locality: (i) $\eps$-locality. (ii)
       $\eps$-hippodrome locality, and %
       (iii) $(\eps,\gamma)$-locality, which is the hippodrome
       locality together with an additional gap requirement. (The
       orange region is the set of points in between $\p$ and $\q$ in
       the ordering.) }
    \figlab{locality}
\end{figure}

This $\eps$-hippodrome locality is sufficient for some applications,
but others require a stronger property of having a ``gap'' in the
ordering.

\begin{defn}
    \deflab{local:gap}%
    For $\gamma \in (0,1/4)$, with $\gamma \geq \eps$ (e.g.,
    $\gamma = 1/4$), an order $\order$ is
    \emphi{$(\eps,\gamma)$-local}, for $\p$ and $q$, if
    \begin{equation*}
        \order(\p,\q)
        \quad\subseteq\quad
        \hippoY{\p\q}{\eps \ell}
        \cap
        \pth{\Bigl.
           \ballY{\p}{\gamma \ell} \cup
           \ballY{\q}{\gamma \ell} },
        \qquad\text{where} \qquad%
        \ell = \dY{\p}{q},
    \end{equation*}
    see \figref{locality} (iii). Namely, the two sets
    $X = \order(\p,\q) \cap \hippoY{\p\q}{\eps \ell} \cap
    \ballY{\p}{\gamma \ell}$ and
    $Y = \order(\p,\q) \cap \hippoY{\p\q}{\eps \ell} \cap
    \ballY{\p}{\gamma \ell}$ have the \emphi{gap property} that
    $\dsY{X}{Y} \geq (1-2\gamma) \ell$.
\end{defn}

\myparagraph{Computation model} %
The model of computation used here is a unit-cost real RAM, supporting
standard arithmetic operations and comparisons (but with the floor
function), augmented with standard bitwise-logical operations
(bitwise-exclusive-or and bitwise-and), which are widely available as
assembly commands on all modern general purpose CPU{}s, and
programming languages. This computational model is reasonable, and is
used (for example) in working with compressed quadtrees efficiently
\cite{h-gaa-11}.

\myparagraph{Previous related work} %
\LSO{}s were introduced by Chan \etal \cite{chj-lota-20}.  Variations
of \LSO{}s for doubling metrics, and other metric spaces, were studied
by Filtser and Le \cite{fl-loars-22} (Triangle \LSO{}, Left-sided
\LSO{}), and Filtser \cite{f-label-23} (Triangle \LSO{}, Rooted
\LSO{}).  Since these works study \LSO in the non-euclidean settings,
they are not directly related to our work here (and they yield,
naturally, worse bounds).  \LSO{}s were used to construct reliable
spanners by Buchin \etal \cite{bho-sda-20, bho-srsal-22} in a simple
plug \& play fashion.  There is some work on quadtrees for hyperbolic
space by Kisfaludi-Bak and van Wordragen \cite{skg-quadt-24}, which
potentially can lead for \LSO{}s for such spaces.

\subsection*{Our results}

We show that there is a universal set of \LSO{}s of $\UC$ that is of
size $O( \Eps^{d-1} \log \Eps)$, such that these orderings have the
somewhat weaker $(\eps,\gamma)$-locality property, where
$\gamma \geq \eps$ is a large fixed constant. This improves the result
of Chan \etal by a factor of $\Eps = 1/\eps$. Furthermore, for many of
the applications using \LSO{}s, one can replace the \LSO{}s by the new
set of ``weaker'' orderings, thus strictly improving their dependence
on $\eps$. In particular, for spanners the new construction enables
maintaining dynamic spanners. Notably, up to $\log \Eps$ factor, the
new construction matches the known lower bound on the number of edges
in a spanner, in the \emph{static} case, thus implying that the new
orderings are almost optimal, and no approach based on \LSO{}s can do
(significantly) better in the Euclidean case.

\myparagraph{Applications} %
Since we have to slightly weaken the definition of \LSO{}s to get the
improved size, we have to rederive the proofs of correctness for some
of the applications. \LSO{}s are especially useful in applications
where one wants to solve some geometric problem under insertions and
deletions of points. Since \LSO{}s decouple the geometric problem from
the data-structure problem (i.e., all one needs to do is to maintain
sorted lists of points under insertions and deletions), this results
in simple data-structures that supports updates:
\begin{compactenumA}
    \smallskip%
    \item \textsf{Locality graph}. In \secref{locality} we outline the
    basic approach -- one maintains a graph over the point set, where
    two points are connected by an edge, if they are adjacent in one
    of the \LSO{}s. The locality graph can be updated in
    $O( \Eps^{d-1} \log^2 \Eps \cdot \log n)$ time, and has the
    property that every point (i.e., vertex) has degree at most
    $O(\Eps^{d-1} \log \Eps)$ in the graph.

    \medskip%
    \item \textsf{Bichromatic closest pair}.  In \secref{bichromatic},
    we show how to use \LSO{}s to dynamically maintain the bichromatic
    closest pair, as the shortest bichromatic edge in the locality
    graph is the desired approximation.
    This also readily leads to a data-structure that supports
    $(1+\eps)$-approximate nearest-neighbor queries. We do not present
    the details for this here, as these results are, relatively
    speaking, less interesting.

    \medskip%
    \item \textsf{Maintaining $(1+\eps)$-spanners.}
    It turns out that the above locality graph is a
    $(1+\eps)$-spanner. Since we can update the locality graph
    quickly, as detailed above, this readily leads to improved dynamic
    data-structures for maintaining spanners.  This is described in
    \secref{spanners}. The spanner construction uses the gap property
    mentioned above.

\end{compactenumA}
\smallskip%
All these results improve by a factor of $\Eps$ over the previous
results of Chan \etal \cite{chj-lota-20}. The new \LSO{}s also
improves, by a factor of $\Eps$, in a plug and play fashion other
results, but the improvement is somewhat less interesting in these
cases (i.e., the improvement is intuitively clear, and in some cases
requires a lot of tedium). This includes reliable spanners
\cite{bho-sda-20}, dynamic approximate minimum spanning trees
\cite{chj-lota-20}, and vertex/edge fault-tolerant spanners
\cite{chj-lota-20}.

\myparagraph{Sketch of the old construction} %
We describe shortly the construction of \LSO{}s by Chan \etal
\cite{chj-lota-20}, see \secref{sketch} for more details.  Chat \etal
reduced the problem of computing \LSOs to the following problem --
consider the integer grid $B= \IRX{4\Eps}^d$, where $\Eps = 1/\eps$
and $\IRX{n} = \{ 1,\ldots, n\}$.  An edge of $\G$ is \emphw{long} if
it corresponds to a pair of points of $B$ that are in distance at
least (say) $2/\eps$ from each other.  The long edges corresponds to
well-separated pairs of points, and these are the ones we must
``serve''. (Understanding why we concentrate on the long edges
requires a bit deeper dive into the details, see \secref{sketch}.)  To
this end, consider the complete graph $\G$ over $B$.  Constructing
\LSOs is then achieved by constructing \LSOs for $B$. A single
ordering (i.e., \LSO) for $B$ is simply a Hamiltonian path of $\G$,
and the task at hand is to compute a small number of Hamiltonian paths
of $\G$ that visits all the long edges. Specifically, a Hamiltonian
path (in a graph) induces an ordering of the vertices, thus
Hamiltonian paths serves as a way to encode the desired
orderings. Having a long edge in such a path, ensures this pair is
consecutive in the associated ordering.

Walecki \cite{a-wwc-08} showed a decomposition of the clique into
Hamiltonian paths, and this leads to the desired \LSOs of $B$. The
problem with this approach is that it requires $\Omega(\Eps^d)$
paths/\LSOs. There is an additional blowup by a factor of
$O( \log \Eps)$ in the number of \LSOs, when converting the \LSOs of
$B$ to \LSOs of $\UC$.

\myparagraph{Sketch of new approach} %
Not all hope is lost however -- the previous construction effectively
ignores the geometry. In addition, one can weaken the requirement, so
that the new \LSOs for $B$ have the $\eps$-hippodrome locality
condition, see \figref{locality} (ii).  So consider two points of $B$
that are far from each other, and the hippodrome induced by the
segment connecting them, see \figref{hip}. We now order the points
inside this region in their order along the direction induced by the
two points. Clearly, this path will have all the long pairs inside the
hippodrome covered (by the transitive closure of this directed
path). The question is how to find such paths, and glue them together
into global orderings that cover all such long edges.

\begin{figure}
    \phantom{}\hfill%
    \includegraphics[page=1]{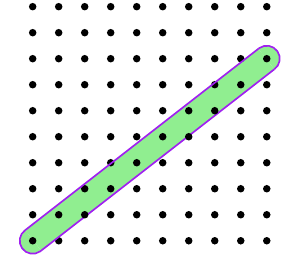}%
    \hfill%
    \includegraphics[page=2]{figs/path}%
    \hfill%
    \phantom{}%

    \caption{A path inside along a hippodrome can cover many long
       edges.} %
    \figlab{hip}
\end{figure}

To this end, we cover the set of directions by an $\eps$-packing of
size $O(\Eps^{d-1})$ (this is where the savings come from!). For each
direction, we show how to construct a constant number of paths that
cover all the long edges that their direction is $\eps$-close to the
current direction. To this end, we compute a packing of the projected
points, and then use coloring to decompose the packing into $O(1)$
packings, such that the points picked in each set, form independent
paths that do not interact with each other. We then take each such set
of paths, and concatenate these paths together to get one
ordering. Overall, we get $O(\Eps^{d-1})$ orderings of $B$, that cover
all the long edges, as desired. Plugging this into the machinery of
Chan \etal then yields the desired \LSOs of $\UC$.

We are hiding some of the technical details under the hood -- for
example, we are dealing with grid cells, and not grid points, etc.
See \secref{constructions} for details of the construction.  The final
result is stated in \thmref{conclusion}.

\myparagraph{Orderings with a gap} %
As mentioned above, a somewhat stronger property of having a gap in
the \LSOs is needed, so that one can construct spanners using
them. The idea is to build \LSO{}s for $B$, by first treating it as a
crude grid, using the previous construction of Chan \etal at the top
grid, and then combine it with the new construction for the bottom
part of the grid. The details are somewhat technical, see \secref{gap}
for details.

\bigskip%
\noindent%
\textbf{Why does the size of \LSOs matter?} %
\LSOs seems to be algorithmically equivalent to other geometric tools,
such as well-separated pair decomposition \cite{ck-dmpsa-95}. Thus,
they seem to be a fundamental property of the underlying space. It is
thus natural to quantify exactly what the minimum size of \LSOs needed
for $\Re^d$, as a function of $d$ and $\eps$. Our work thus can be
interpreted as (almost) settling this question.

\subsection*{Paper organization}

We describe the old construction of Chan \etal in \secref{sketch}. We
present some necessary standard results on nets/packings in
\secref{packing}.  In \secref{projections}, we describe how to
construct the needed packings on projected points, and how to
decompose such a packing into a packing of larger radius.  At last, we
are ready. We present, in \secref{basic}, the construction of the
orderings with the hippodrome property, and, in \secref{gap}, we
present the modified construction with the gap property.

The applications are presented in \secref{applications}.  In
\secref{locality} we show how the orderings give rise to a locality
graph that captures all the information needed for the
applications. In particular, this graph can be maintained efficiently
under insertions and deletions.  In \secref{bichromatic}, we present
the improved dynamic data-structure for dynamic bichromatic closest
pair.  In \secref{spanners}, we present the improved result for
spanners. We also provide a proof of correctness showing that the new
orderings are sufficient to guarantee that the locality graph is a
spanner. We also point out, in \secref{lower:bounds}, that any \LSOs
that their locality graph is a spanner, must have $\Omega(\Eps^{d-1})$
orderings, thus implying that our construction of \LSOs is tight up to
a factor of $\log \Eps$.

\section{Preliminaries}
\subsection{Definitions}

\begin{defn}
    For two sets $X, Y \subseteq \Re^d$, let
    $\dsY{X}{Y} = \min_{x \in X, y\in Y} \dY{x}{y}$ denote the
    \emphi{distance} between $X$ and $Y$. For a point $\p \in \Re^d$,
    we use the shorthand $\dsY{p}{X} = \dsY{\{p\}}{X}$.
\end{defn}

\subsection{Locality sensitive orderings}
\seclab{sketch}

For a set $X$, consider a total order (or \emphw{ordering}) $\prec$ on
the elements of $X$.  Two elements $x,y \in X$ are \emphi{adjacent} if
there is no element $z \in X$, such that $x \prec z \prec y$ or
$y \prec z \prec x$. Our purpose is to define a small number of
orderings on $\UC$ so that certain geometric tasks can be carried out
using these orderings.

In the following, we fix a parameter $\eps = 1/2^\LgEps$, for some
integer $\LgEps >0$, and use
\begin{equation}
    \Eps = 1/\eps=2^\LgEps
    \qquad\text{and}\qquad %
    \LgEps = \log_2 \Eps.
    \eqlab{eps}
\end{equation}

Our starting point is the result of Chan \etal \cite{chj-lota-20}.
Since we use the same framework, we quickly sketch the above
construction (reproducing some relevant definitions).

\begin{defn}
    \deflab{t:grid}%
    Let $\Cube \subseteq \Re^d $ be an axis-parallel cube with side
    length $\num$.  For an integer $t > 1$, partitioning $\Cube$
    uniformly into a $t \times t \times \cdots \times t$ subcubes,
    forms a \emphi{$t$-grid} $\Grid(\Cube, t)$.  The diameter of a
    cube $\Cell$ is
    $\diamX{\Cell} = \mathrm{sidelength}(\Cell)\sqrt{d}$, and let
    $\cenX{\Cell}$ denote its \emphi{center}. Let $\ballX{\Cell}$ be
    the smallest ball enclosing $\Cell$ -- that is,
    $\ballX{\Cell} = \ballY{\cenX{\Cell}}{{diam(\Cell)}/{2}}$
\end{defn}

\begin{defn}
    An \emphi{$\eps$-quadtree} $\qte$ is a quadtree-like structure,
    built on a cube with side length $\num$, where each cell is
    partitioned into an $\Eps$-grid.  The construction then continues
    recursively into each grid cell of interest.  As such, a node in
    this tree has up to $\Eps^d$ children, and a node at level
    $i \geq 0$ has an associated cube of side length $\num
    \eps^i$. When $\Eps = 2$, this is a regular quadtree.
\end{defn}

Informally, an $\eps$-quadtree connects a node directly to its
decedents that are $\LgEps$ levels below it in the regular
quadtree. Thus, to cover the quadtree, we need $\LgEps$ different
$\eps$-quadtrees to cover all the levels of the quadtree. This can be
achieved by setting directly the level of the root of the
$\eps$-quadtree. Specifically, we fix the root of the $\eps$-quadtree
$\qte^i$ to be $[0,2^{i})^d$, for $i=1, \ldots, \LgEps$. This implies
the following.
\begin{lemma}
    \lemlab{reg:to:eps:quad}%
    Let $\qt$ be a regular (infinite) quadtree over $[0, 2)^d$.  There
    are $\LgEps$ $\eps$-quadtrees $\qte^1, \ldots, \qte^{\LgEps}$,
    such that the collection of cells at each level in $\qt$ is
    contained in exactly one of these $\eps$-quadtrees, where
    $\LgEps = \log_2 \Eps$, see \Eqref{eps}.
\end{lemma}

The problem with using a quadtree, for our purposes, is the alignment
of the boundaries of their cells. Intuitively, two points that are
close together might be separated by quadtree boundary, that belongs
to cells that are dramatically bigger than the distance between the
points. Fortunately, this problem can be overcome by shifting (the
point set, or the quadtree) $d+1$ times.

\begin{lemma}[\tcite{c-annqr-98},~Lemma~3.3]
    \lemlab{shifting}%
    Consider any two points $\p, \q \in \UC$, and let $\qt$ be the
    infinite quadtree of $[0,2)^d$.  For $D= 2\ceil{d/2}$ and
    $i = 0, \ldots, D$, let $\shift_i = (i/(D+1), \ldots,
    i/(D+1))$. Then there exists an $i \in \brc{0, \ldots, D}$, such
    that $\p + \shift_i$ and $\q + \shift_i$ are contained in a cell
    of $\qt$ with side length $\leq 2(D+1) \dY{\p}{\q}$.%
\end{lemma}

The final tool needed is a small set of orderings, such that any two
subcells, in a $t$-grid, are adjacent in one of the orderings.

\begin{lemma}[\tcite{chj-lota-20}]
    \lemlab{all:friends}%
    Let $S$ be set of $t^d$ subcells of a $t$-grid
    $\Grid(\Cube, t)$.
    There is a set $\Pi$  of $O(t^d)$ orderings of
    $S$, such that each pair of elements of $S$ is adjacent in at
    least one of the orderings.
\end{lemma}
\begin{proof}
    The original proof is a direct consequence of a beautiful result
    of Walecki from the $19$\th century \cite{a-wwc-08}, see
    \cite{chj-lota-20} for details. For the sake of completeness, we
    sketch a shorter and inferior argument -- a random permutation
    (i.e., order) of $S$ has a specific pair being consecutive with
    probability at least $1/\cardin{S} = 1/t^d$. Repeating this
    process $m = O(t^d \log t)$ times, implies that the probability
    that there is any pair of elements in $S$ that is not consecutive,
    in one of the orderings, is
    $\leq \cardin{S}^2 (1-1/t^d)^m \ll 1/2$.  Thus, the computed set
    of orderings has the desired property with constant probability.
\end{proof}%

\subsubsection{From order on a grid, to orderings of the unit cube}
\seclab{order:construction}

Chan \etal \cite{chj-lota-20} showed how to convert an order of a
$t$-grid into an ordering of $[0,1)^d$. We now describe this
construction in detail.  Let $\Forest$ be the set of $\LgEps$
$\eps$-quadtrees of \lemref{reg:to:eps:quad}, and let
$\Shifts \subseteq \UC$ be the set of at most $d+1$ shifts of
\lemref{shifting}.  Let $\OYC$ be the set of orderings of the $t$-grid
$\Grid(\Cube, \Eps)$, for some cube $\Cube$, as constructed by
\lemref{all:friends}.

Next, consider an $\eps$-quadtree $\qte \in \Forest$, a shift
$\shift \in \Shifts$, and an ordering $\order \in \OYC$. This
immediately induces an ordering over the points of $[0,1)^d$. Indeed,
consider two points $\p,\q \in [0,1)^d$, and consider their shifted
image $\p' = \shift + \p$ and $\q' = \shift+\q$. There is a unique
least common ancestor (\LCA) node $u \in \qte$ to the leafs where
$\p'$ and $\q'$ would be stored in the quadtree. The node $u$ has
$\Eps^d$ children, with $\p'$ and $\q'$ belonging to two different
children, say $u_\p$ and $u_\q$. If $\order$ has $u_\p$ before $u_\q$,
then we consider $\p \prec_\order \q$, and otherwise
$\p \succ_\order \q$.  Using standard bit operations, and the ordering
$\order$ listed explicitly (which requires $O(\Eps^d)$ space), this
can be done in $O( \LgEps) = O( \log \tfrac{1}{\eps})$ time
\cite{chj-lota-20}. (The quadtree $\qte$ is used implicitly, so it is
not computed explicitly.)  Overall, the number of orderings this
yield, is $\Of( \cardin{\OYC} \cardin{\Shifts} \LgEps )$. Given two
points comparing their order according to a specific given order takes
$O( \LgEps)$ time. The resulting set of orderings of $[0,1)^d$ is of
size $O( \Eps^d \log \Eps)$.  This implies the following theorem, see
\cite{chj-lota-20} for more details.

\begin{theorem}[\tcite{chj-lota-20}]
    \thmlab{lso:original}%
    There is a set $\ordAll$ of $\Of( \Eps^d \log \Eps)$ orderings of
    $[0,1)^d$, such that for any two points $\p, \q \in [0,1)^d$ there
    is an ordering $\order \in \ordAll$ defined over $[0,1)^d$, such
    that for any point $\pC$, with
    $\p \precX{\order} \pC \precX{\order} \q$, we have that either
    $\dY{\p}{\pC} \leq \eps \dY{\p}{\q}$ or
    $\dY{\q}{\pC} \leq \eps \dY{\p}{\q}$.
\end{theorem}

\subsection{Coverings and packings}
\seclab{packing}

\begin{defn}
    A unit length vector $v \in \Re^d$ is a \emphi{direction}. The set
    of all directions form the \emphi{unit sphere} $\Sphere$ centered
    at the origin.
\end{defn}

\begin{defn}
    \deflab{packing}%
    Consider a set $\P \subseteq \Re^d$, and a parameter $\rad$.  A
    set $\Packing \subseteq \P$ is an \emphi{$\rad$-covering} of $\P$,
    if for all $\p \in \P$, there exists a point $\p' \in \Packing$,
    such that $\dY{\p}{\p'} \leq \rad$.  A point set $\N$ is
    \emphi{$\rad$-separated} if for any two distinct points
    $\p, \q \in \N$, we have $\dY{\p}{\q} > \rad$.  A subset
    $\N \subseteq \P$ that is both $\rad$-separated and an
    $\rad$-covering of $\P$ is an
    \emphi{$\rad$-packing}\footnote{Confusingly $r$-packings are also
       known as $r$-nets, but as the $r$-net is an overloaded concept,
       we use the alternative, hopefully less-confusing, term.} of
    $\P$.
\end{defn}

For a finite set $\P$ in $\Re^d$, an $\rad$-packing can be computed in
linear time \cite{hr-nplta-15} (for constant $d$). We need such a
packing of the unit sphere.

\begin{lemma}
    \lemlab{packing}%
    For any $\rad \in (0,1)$, there is an $\rad$-packing $\Packing$ of
     the unit sphere $\Sphere$ of size $O( 1/\rad^{d-1})$, and it can be
    computed in $O(1/\rad^{d-1})$ time.
\end{lemma}

\begin{proof}
    An $\rad$-covering of the sphere of the desired size follows by
    taking the uniform grid in $\Re^d$ of sidelength
    $\rad/\sqrt{d}$. Let $\Cells$ be the set of all grid cells that
    intersect the boundary of the unit sphere $\Sphere$. Clearly,
    $\cardin{\Cells} = O(1/\rad^{d-1})$. We pick an arbitrary point on
    the boundary of the sphere inside each such cell, and let $N'$ be
    the resulting set of points. Since the diameter of each grid cell
    is at most $\rad$, it follows that $N'$ is an $\rad$-covering.

    The set $N'$ has (asymptotically) the right
    size. Converting it into an $\rad$-packing can be done in linear
    time using the algorithm mentioned above \cite{hr-nplta-15}.
\end{proof}

\begin{remark}
    \remlab{packing:l:b}%
    A standard packing argument implies that any $\eps$-packing of the
    unit sphere in $\Re^d$ must be of size $\Theta( \Eps^{d-1})$.
\end{remark}

\section{The new construction of \LSOs}
\seclab{constructions}

\subsection{Projections}
\seclab{projections}

\begin{defn}
    For a direction $v$, the \emphi{projection} by $v$, of a point
    $p \in \Re^d$, denoted by $\projY{v}{p} = p - v \DotProdY{p}{v}$,
    is the projection of $p$ onto the hyperplane passing through the
    origin and perpendicular to $v$, where $\DotProdY{p}{v}$ denotes
    the dot-product of $p$ and $v$.  For two points $\p,\q \in \Re^d$,
    let $\dZ{v}{\p}{\q} = \dY{\projY{v}{\p}}{ \projY{v}{\q}}$ be the
    distance between the two projected points.
\end{defn}

\begin{lemma}
    \lemlab{intersection_nonempty}%
    Let $R > \tau > 0$ be parameters. One can compute, in
    $O(\cardin{\DSet}) = O\bigl( (R/\tau)^{d-1} \bigr)$ time, a set
    $\DSet$ of directions. We have that, for all $\p,\q \in \Re^d$,
    with $\dY{\p}{\q} \leq R$, there exists a direction $v \in \DSet$,
    such that $\dZ{v}{\p}{\q} \leq \tau$.
\end{lemma}
\begin{proof}
    Let $\DSet$ be an $\epsA$-packing of the unit sphere $\Sphere$
    computed by \lemref{packing}, where $\epsA = \min(\tau/R,1/4)$. We
    have
    $\cardin{\DSet} = O(1/\epsA^{d-1}) = O\bigl( (R/\tau)^{d-1}
    \bigr)$.  As for correctness, consider the line $\Line$ connecting
    $\p$ to $\q$. Let $u$ be the direction vector of $\Line$, and let
    $v \in \DSet$ be the closest direction to $u$.  By construction,
    $\dY{u}{v} \leq \epsA$. Let $\LineA$ be the line passing through
    $\p$ in the direction of $v$, and let $\q'$ be the projection of
    $\q$ to this line, see \figref{angles}.

    \begin{figure}[h]
        \phantom{}%
        \hfill%
        {\includegraphics[page=1]{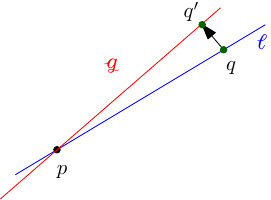}} \hfill%
        \phantom{}%

        \caption{Illustration of the proof of
           \lemref{intersection_nonempty}.}%
        \figlab{angles}
    \end{figure}
    Clearly,
    \begin{math}
        \dZ{v}{\p}{\q} =%
        \dY{\q}{\q'} =%
        \dY{\p}{\q} \sin \angle \q \p \q' \leq R \sin \epsA \leq R
        \epsA \leq R\frac{\tau}{R} = \tau,
    \end{math}
    since $\dY{\p}{\q} \leq R$, $\sin x \leq x$ for $x$ positive,
    $\epsA > 0$, and $\epsA \leq \tau/R$.
\end{proof}

\begin{lemma}
    \lemlab{spearated}%
    Let $\N \subseteq \Re^d$ be an $r$-packing of some set in $\Re^d$,
    for some real number $r$. For a parameter $R > r$, the set $\N$
    can be partitioned, into $m=O\bigl( (R/r)^d)$ disjoint sets
    $\N_1, \ldots, \N_m$, such that, for all $i$, $\N_i$ is
    $R$-separated. The partition can be computed in $O( n m)$ time,
    where $n = \cardin{\N}$.
\end{lemma}
\begin{proof}
    The algorithm computes in each iteration an $R$-packing of $\N$,
    removes its points from $\N$, and repeat the process till $\N$ is
    exhausted. The bound on the number of resulting sets follows by an
    easy packing argument. Indeed, if a point $\p \in \N$ is removed
    in the $i$\th iteration, then it is in distance $\leq R$ from all
    the previous point sets extracted (as otherwise, the earlier
    computed point-sets would not have been $R$-packings). However,
    the original point set $\N$ is $r$-separated, which implies that
    this can happen at most $O\bigl( (R/r)^d \bigr)$ times. Since
    computing a net can be done in linear time, the claim follows.
\end{proof}

\subsection{Constructing the ordering}
\seclab{basic}

\begin{lemma}
    \lemlab{grid_conclusion} %
    Consider a \defrefY{$t$-grid}{t:grid} $\Grid = \Grid(\Cube, t)$ of
    an axis-parallel cube $\Cube \subseteq \Re^d $, where $t$ is a
    fixed positive integer. Then, one can compute a set
    $\OYC = \OY{t}{d}$ of $O(t^{d-1})$ orderings of the cells of
    $\Grid$, such that for any two cells $\Cell_1, \Cell_2 \in \Grid$,
    there exists an ordering $\sigma \in \OYC$, where for all
    $\Cell_1 \precX{\sigma} \Cell_3 \preceqX{\sigma}\Cell_4
    \precX{\sigma} \Cell_2$, we have:
    \begin{compactenumi}
        \smallskip%
        \item
        $\dY{\cenX{\Cell_3}}{\cenX{\Cell_4}} \leq
        \dY{\cenX{\Cell_1}}{\cenX{\Cell_2}}$,

        \medskip%
        \item $\Cell_3$ and $\Cell_4$ each intersects the segment
        $\cenX{\Cell_1}\cenX{\Cell_2}$.
    \end{compactenumi}
    \smallskip%
    Here $\cenX{\Cell}$ denotes the center point of $\Cell$.
\end{lemma}

\begin{figure}[h]
    \phantom{}%
    \hfill%
    \includegraphics[page=2]{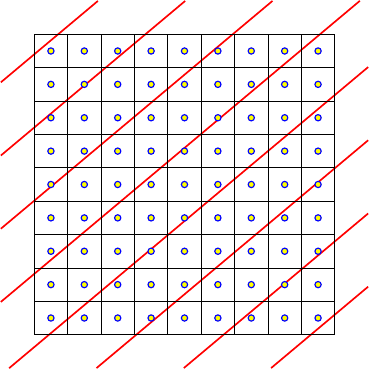} %
    \hfill%
    \includegraphics[page=32]{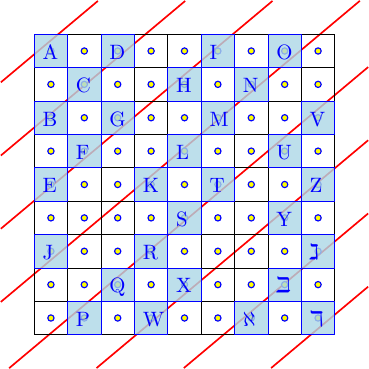} %
    \hfill%
    \phantom{}%
    \caption{Illustration of one of the orderings of
       \lemref{grid_conclusion}.}
    \figlab{animation}
\end{figure}

\begin{proof}
    Let $\num$ be the sidelength of the individual cells of $\Grid$.
    Thus $\sdX{\Cube} = t \num$. Let
    $\P = \Set{\cenX{\Cell}}{\Cell \in \Grid(\Cube, t)}$,
    $R = \diamX{\Cube} = t \sqrt{d} \num$. and let $\tau =
    \num/(8d)$. Let $\DSet$ be the set of directions as computed by
    the algorithm of \lemref{intersection_nonempty} for $\tau/2$ and
    $R$. Observe that $R/(\tau/2) = O(t)$, and
    $\cardin{\DSet} = O(t^{d-1})$.

    For each direction $v \in \DSet$, consider the projected cube
    $\Cube_v = \projY{v}{\Cube}$, the projected point set
    $\P_v = \projY{v}{\P}$, and let $\N_v$ be a $\tau/2$-packing of
    $\P_v$ (which is a point set lying on the hyperplane perpendicular
    to $v$ passing through the origin).  Using \lemref{spearated},
    split $\N_v$ into $\psi$-separated sets $\N_v^1, \ldots, \N_v^m$,
    where $\psi = 2 \sqrt{d} \num$, and
    $m = O\bigl( (\psi/\tau)^{d-1} \bigr) = O(1)$ sets. Observe that
    for any $\Cell \in \Grid$, its projection $\projY{v}{\Cell}$ has
    diameter $\leq \sqrt{d} \num < \psi\Bigr.$. Namely, no projected
    cell can be stabbed by two points belonging to the same
    $\psi$-separated set $\N_v^i$.  The algorithm computes an ordering
    for each set $N_v^i$.  The basic idea is to start with an empty
    ordering, and then for each point $\p \in \N_v^i$, the algorithm
    would concatenate (say, to the end of the ordering computed so
    far) some of the projected cells that $\p$ stabs (in their order
    along the line induced by $\p$). All the remaining unassigned
    cells would be concatenated to the resulting order in the end, in
    an arbitrary fashion.

    We now describe this in more detail. Fixing the values of $v$ and
    $i$, identifies a set $\N_v^i$. For
    $j=1,\ldots, \cardin{\N_v^i }$, let $p_j$ be a point of $\N_v^i$
    not handled yet. The algorithm computes the oriented line
    $\Line_j$ passing through $p_j$ in the direction of $v$.  Let
    $Z_j$ be the set of all the cells of $\Grid$ that $\Line_j$ stabs,
    and their center is in distance at most $\tau$ from $\Line_j$. All
    such cells intersect $\Line$ in a ``long'' interval of length
    $\geq \num - 2\tau \geq (3/4)\num \geq 6 \tau$ (that is, all the
    cells of $Z_j$ have their centers ``almost'' on $\Line_j$). Thus,
    the distance between the centers of the cells of $Z_j$ in the
    original space, or the distance between their corresponding
    projections on $\Line_j$ are the same up to $\pm 2\tau$.  In
    particular, we sort the cells of $Z_j$ according to the order of
    their projected centers along $\Line_j$, and append them (in this
    sorted order) to the computed order.  This process is illustrated
    in \figref{animation}.  Note that in the end of this process, most
    grid cells of $\Grid$ are not included in the computed order. We
    append all these yet unordered cells in an arbitrary fashion to
    the end of the ordering.  Let $\sigma(v,i)$ denote the resulting
    ordering of the cells of $\Grid$.

    Finally, let
    \begin{equation*}
        \OY{t}{d}=\Set{ \sigma(v,i)}{v \in \DSet, i =1,\ldots, m}
    \end{equation*}
    denote the resulting set of orderings.

    \medskip

    As for correctness, consider any two cells
    $\Cell_1, \Cell_2 \in \Grid$. By construction there is a direction
    $v \in \DSet$, such that the projected centers
    $\cc_1 = \projY{v}{\cenX{\Cell_1}}$ and
    $\cc_2 = \projY{v}{\cenX{\Cell_2}}$ are in distance at most
    $\tau/2$ from each other. In particular, let $\p$ be the closest
    point to either of them in the $\tau/2$-net $\N_v$. Clearly, $\p$
    is in distance at most $\tau$ from $\cc_1$ and from $\cc_2$. In
    particular, let $i$ be the index, such that $\p \in \N_v^i$, and
    observe that when the algorithm handles $\p$, say in the $j$\th
    inner iteration (so $\p = \p_j$), it would include both $\Cell_1$
    and $\Cell_2$ in the ordering (as they both belong to the set
    $Z_j$), and all the cells included in the ordering in between
    these two cells are in $Z_j$ and lie in between these two cells
    along $\Line_j$. This readily implies properties (i) and (ii).
\end{proof}

\subsection{An ordering with a gap}
\seclab{gap}

For a segment $\seg$ and a constant $c$, let $c \seg$ denote the
scaling of $\seg$ by a factor of $c$ around its center. We require the
following additional property: If any two cells in a grid are far
enough from each other, then there is an ordering, such that between
the two cells in the ordering there is a ``gap'' which is roughly the
distance between the two cells. This is essentially a weakened version
of the property that the original locality sensitive orderings had.

\begin{lemma}
    \lemlab{grid:order:gap} %
    Consider a $t$-grid $\Grid = \Grid(\Cube, t)$ of an axis-parallel
    cube $\Cube \subseteq \Re^d $ with the cells having sidelength
    $\num$, and a parameter $\alpha \in \IRX{ t} = \{1,\ldots, t\}$,
    where $t$ is some fixed positive integer that is a power of
    two. Then, one can compute a set $\OYC = \OZ{t}{\alpha}{d}$ of
    $O(t^{2d-1}/\alpha^d)$ orderings of the cells of $\Grid$, such
    that for any two cells $\Cell_1, \Cell_2 \in \Grid$, with
    $\dsY{\Cell_1} {\Cell_2} \geq \alpha \num$, there exists an
    ordering $\sigma \in \OYC$, such that for all $\Cell \in \Grid$,
    with
    \begin{math}
        \Cell_1 \precX{\sigma} \Cell \precX{\sigma} %
        \Cell_2,
    \end{math}
    we have:
    \begin{compactenumi}
        \smallskip%
        \item $\Cell$ intersects the segment
        $\cenX{\Cell_1}\cenX{\Cell_2}$.

        \smallskip%
        \item $\dsY{\Cell}{\Cell_1} \leq (\alpha/4) \num$ or
        $\dsY{\Cell}{\Cell_2} \leq (\alpha /4) \num$.
    \end{compactenumi}
\end{lemma}
\begin{figure}
    \centerline{\includegraphics{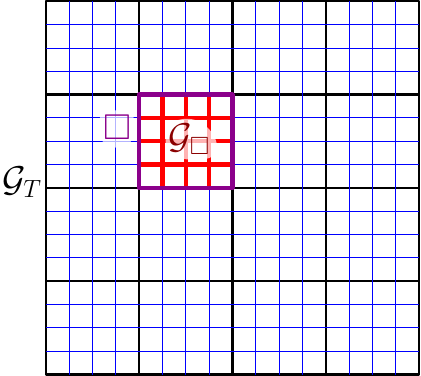}}
    \caption{The large grid and its refinement.}
    \figlab{proof:g:o:g}
\end{figure}
\begin{proof}
    Assume $t=2^i$ for some integer $i$. And let $j$ be the largest
    integer, such that $\beta = 2^j < \ceil{\alpha/8d}$.  We divide
    $\Cube$ into a grid of ``large'' cells
    $\Grid_T = \Grid(\Cube, t/\beta)$. Observe that every cell of
    $\Grid_T$ has diameter
    \begin{equation*}
        \beta \sqrt{d} \num
        <
        \ceil{\alpha/8d}\sqrt{d} \num
        <
        (\alpha/4) \num,
    \end{equation*}
    and
    \begin{math}
        N =%
        \cardin{ \Grid_T} =%
        \Theta\bigl( (t/\beta)^d)%
        =%
        \Theta\bigl( (t/\alpha)^d).
    \end{math}
    We partition every cell $\Cell$ of $\Grid_T$ into a secondary grid
    $\Grid_\Cell = \Grid(\Cell, \beta)$, see
    \figref{proof:g:o:g}. After applying this to all the grid cells of
    $\Grid_T$, this results in the original grid $\Grid$. Let $\OYC_T$
    be the set of orderings, of \lemref{all:friends}, of $\Grid_T$
    that has the property that every two cells of $\Grid_T$ are
    adjacent in one of the orderings, where
    $\cardin{\OYC_T}=O(N)$. Let $\OYC_B$ be the set of orderings for
    the original grid $\Grid$ from \lemref{grid_conclusion}. Consider
    an order $\sigma_T \in \Grid_T$ and an order
    $\sigma_B \in \OYC_B$. They induce an ordering of the cells of
    $\Grid$ as follows -- the top ordering $\order_T$ orders the cells
    of $\Grid$ into secondary blocks, where each secondary block is
    all the cells of $\Grid$ that lie in a single cell of
    $\Grid_T$. Every such block is then sorted using the order
    $\order_B$. This defines a new set of orders $\OYC$ over $\Grid$
    of size
    $O( \cardin{\OYC_T} \cdot \cardin{\OYC_B}) =
    O(t^{2d-1}/\alpha^d)$.

    Consider two cells $\Cell_1, \Cell_2 \in \Grid$, such that
    $\dsY{\Cell_1}{\Cell_2} \geq \alpha \num$. Let
    $\Cell_1', \Cell_2' \in \Grid_T$ be the two cells containing
    $\Cell_1$ and $\Cell_2$, respectively. As $\Cell_1$ and $\Cell_2$
    are ``far'' from each other, the cells $\Cell_1'$ and $\Cell_2'$
    are distinct, and not adjacent in $\Grid_T$. Let $\order_T$ be the
    ordering having $\Cell_1'$ and $\Cell_2'$ adjacent in the
    ordering. By \lemref{grid_conclusion}, there is an order
    $\order_B \in \OYC_B$, such that all the grid cells in between
    $\Cell_1$ and $\Cell_2$ in the ordering stab the segment
    connecting their centers.  The combined ordering
    $(\order_T,\order_B)$ has between $\Cell_1$ and $\Cell_2$ only
    cells that intersect the segment connecting their centers, that
    lie inside $\Cell_1'$ or $\Cell_2'$, as desired.
\end{proof}

\begin{remark}
    The number of orderings in \lemref{grid:order:gap} seems
    excessive, but it is sufficient for our application, as we are
    going to use $\alpha = \Omega(t)$, where the resulting total
    number of orderings is
    \begin{math}
        O(t^{2d-1}/\alpha^d) =%
        O(t^{d-1}).
    \end{math}
\end{remark}

\subsection{Result}

\begin{theorem}
    \thmlab{conclusion}
    Let $\eps, \gamma \in (0, 1/2]$ be fixed constants, such that
    $\gamma \geq \eps$, and let $\Eps=1/\eps$.  Then, there is a set
    $\Pi$ of $ m= O\bigl( (\Eps^{d-1}\log \Eps) / \gamma^{d}\bigr)$
    orderings, of $[0, 1)^d$,such that for any two points
    $\p, \q \in [0, 1)^d$, there is an ordering $\sigma \in \Pi$
    defined over $[0, 1)^d$, such that
    \begin{compactenumi}
        \smallskip%
        \item for all points $u$, with
        $p \precX{\sigma} u \precX{\sigma} q$, we have
        $\dsY{u}{\p\q} \leq \eps \dY{\p}{\q}$ and
        $\dsY{u}{ \{\p, \q\} } \leq \gamma \dY{\p}{\q}$,

        \smallskip%
        \item for all points $u, v$ with
        $p \precX{\sigma} u \precX{\sigma} v \precX{\sigma} q$, we
        have $\dY{u}{v} \leq (1+\eps) \dY{\p}{\q}$.
    \end{compactenumi}
    \smallskip%
    Namely, $\Pi$ is a set of $m$ orderings that satisfy the
    $(\eps,\gamma)$-locality property, see
    \defref{local:gap}.
\end{theorem}
\begin{proof}
    Let $\eps' = 1/2^\LgEps$, for the minimum $\LgEps$ such that
    $ \eps' < \eps/(4d^2)$.  Let
    \begin{equation*}
        \Eps = 1/\eps',
        \qquad\text{and}\qquad%
        \alpha = \max\bigl(1, \floor{\gamma \Eps/(2d^2)} \bigr)
    \end{equation*}
    Let $\OYC = \OZ{\Eps}{\alpha}{d}$ be the set of orderings of
    $\Grid(\Cube, \Eps)$ as defined by \lemref{grid:order:gap}.  As
    described in \secref{order:construction}, the set $\OYC$ induces a
    set $\Pi = \Pi( \eps', \alpha)$ of orderings of $[0,1)^d$, where
    \begin{equation*}
        \cardin{\Pi}%
        =%
        O\Bigl( \frac{\Eps^{2d-1}}{\alpha^d} \LgEps \Bigr)
        =%
        O\Bigl( \frac{\Eps^{d-1}}{\gamma^d} \LgEps \Bigr)
        =%
        O\Bigl( \frac{1}{\eps^{d-1} \gamma^d} \log \frac{1}{\eps} \Bigr).
    \end{equation*}

    So consider any two points $\p,\q \in [0,1)^d$. There is a shift
    $\shift \in \Shifts$, and an $\eps'$-quadtree $\qte \in \Forest$,
    such that the \LCA of $\p$ and $\q$ in $\qte$ is a node $w$, such
    that $\sdX{\Cell_{w} } \leq (d+1)\dY{\p}{\q}$, see
    \lemref{shifting}, where $\sdX{\Cell} = \diamX{\Cell} / \sqrt{d}$
    is the \emphw{sidelength} of $\Cell$.  In particular, let
    $\zeta = \sdX{\Cell_{w}}/\Eps$ be the sidelength of the subcells
    of the grid of $\qte$ for $\Cell_w$.  Observe that
    $ \dY{\p}{\q} \geq \Eps \zeta /(d+1)$.  Let $\Cell_1$ and
    $\Cell_2$ be the cells of the two children of $w$ that contains
    $\p$ and $\q$, respectively. In particular, we have that
    \begin{equation*}
        \dsY{\Cell_1}{\Cell_2}
        \geq%
        \frac{\Eps \zeta}{d+1} - 2\sqrt{d} \zeta
        \geq
        \alpha \zeta.
    \end{equation*}
    Now, \lemref{grid:order:gap} guarantees the existence of an
    ordering $\order \in \OZ{\alpha}{\Eps}{d}$ that stabs $\Cell_1$
    and $\Cell_2$, and all the cells in between in this ordering are
    ``$(\alpha/4)$-close'' to either $\Cell_1$ or $\Cell_2$. In
    particular, the ordering of $[0,1)^d$ induced by $\shift, \qte$
    and $\order$ has the desired property.
\end{proof}

\begin{remark}
    \remlab{operation}%
    Computing the orderings of \thmref{conclusion} can be done in
    $\Eps^{O(d)}$ time, and since this is a preprocessing stage, we
    ignore this running time in the theorem statement.  This results
    in $ m= O\bigl( (\Eps^{d-1}\log \Eps) / \gamma^{d}\bigr)$
    orderings, each order requires $O(\Eps^d)$ space to store it. In
    the following, $\gamma$ is a fixed ``large'' constant (e.g.,
    $\gamma=1/8$), so the overall space to store these orderings is
    $O(\Eps^{2d-1} \log \Eps )$. Given an order $\order$ and two
    points, comparing the two points according to $\order$ requires
    answering a single \LCA query on a quadtree, see
    \secref{order:construction}. With the appropriate bit operations
    such a query can be carried out in $O( \log \Eps)$ time
    \cite{chj-lota-20}.
\end{remark}

\section{Applications}
\seclab{applications}

\subsection{Locality graph}
\seclab{locality}

Given an ordering $\order$ of \thmref{conclusion}, one can maintain a
sorted list of $n$ points in $[0,1)^d$ in $O( \log n \log \Eps)$ time
per insertion/deletion. Indeed, using any standard balanced binary
search tree storing $n$ elements, requires $O( \log n)$ time per
operation, and each such operation performs $O( \log n)$
comparisons. Each comparison takes $O( \log \Eps)$ time to perform,
see \remref{operation}. In particular, a natural approach is to
maintain a dynamic graph, over the (evolving) point set $\P$, where
two points are connected by an edge $\iff$ the two points are adjacent
in one of the orderings of $\Pi$ provided by \thmref{conclusion}. Note
that any insertion/deletion would cause $O( \Eps^{d-1} \log \Eps)$
edges to be inserted/deleted to this locality graph of $\P$.

\begin{defn}
    \deflab{locality:graph}%
    For parameters $\eps, \gamma \in (0,1)$, and a set of points
    $\P \subseteq \UC$, let $\GL = \GL(\P, \eps, \gamma)$ be the
    \emphi{locality graph}, described above, for the set of orderings
    $\Pi$ computed by \thmref{conclusion}.
\end{defn}

\newcommand{\cEdges}{\Mh{\Psi}}%
\begin{lemma}
    For $\eps \in (0,1)$ and $\gamma \in (1/32,1)$ (with
    $\gamma \geq \eps$), the locality graph
    $\GL = \GL(\P, \eps,\gamma)$ defined for a set $\P$ of $n$ points
    in $\UC$, has $O( \cEdges n )$ edges, where
    $\cEdges =O( \Eps^{d-1} \log \Eps)$.  An insertion/deletion into
    $\P$ can be performed in
    $O\bigl( (\cEdges \log \Eps) \log n \bigr)$ time, and involve the
    deletion/insertion of at most $\cEdges$ edges of $\GL$.
\end{lemma}

\subsection{Bichromatic closest pair}
\seclab{bichromatic}

\begin{theorem}
    \thmlab{bichromatic_closet_pair}%
    Given two sets of points, $R$ and $B$, both subsets of $[0, 1)^d$,
    and a parameter $\eps \in (0, 1/2)$, one can maintain a
    $(1 + \eps)$-approximation to the closest bichromatic pair
    $R \times B$.  Each insertion/deletion takes
    $O( \Eps^{d-1} \log^2 \Eps \cdot \log {n} )$ time per operation,
    where $n$ is the maximum size of $|R| + |B|$ over time, and
    $\Eps = 1/\eps$. This data-structure use $O( \cEdges n )$ space,
    where $\cEdges =O( \Eps^{d-1} \log \Eps)$. At all times, it
    maintains a pair of points $r \in R$ and $b \in B$, such that
    $\dY{r}{b} \leq (1 + \eps) \dsY{R}{B}$, where
    $\dsY{R}{B} = \min_{b \in B, r \in R} \dY{b}{r}$.
\end{theorem}
\begin{proof}
    The algorithm maintains the locality graph
    $\GL = \GL(\P, \eps/4, 1/8)$. Furthermore, it maintains all the
    bichromatic edges of $\GL$ in a min-heap sorted by their
    length. We claim that the minimum of this heap is the desired
    approximation. The bounds stated readily follows from the above
    discussion.

    Consider the bichromatic closest pair $r \in R$ and $b \in B$. By
    \thmref{conclusion}, there exists an order $\order \in \Pi$, such
    that
    \begin{equation*}
        \forall \p \in R\cup B:  r \precX{\order} \p \precX{\order} b
        \qquad\implies
        \qquad
        \dsY{\p}{rb} = \min_{\q \in rb} \dY{\q}{\p} \leq
        (\eps/4)\dY{r}{b}.
    \end{equation*}
    This implies that any pair of points of $\P$ in the (close)
    interval $\order[r,b]$, see \Eqref{interval}, must be in distance
    at most $(1+\eps/2)\dY{r}{b}$ from each other, as this is the
    diameter of the hippodrome $\hippoY{rb}{\eps \dY{r}{b}}$. So
    consider the sorted list of $\P \cap \order[r,b]$ (according to
    $\order$), and observe that it starts in a point of $R$ and ends
    in a point of $B$. As such, there must be a consecutive
    bichromatic pair, and its this appear as an edge in the locality
    graph. This candidate pair already provides the desired
    approximation, and the data-structure can only (potentially)
    return a better one.
\end{proof}%

\subsection{Dynamic spanners}
\seclab{spanners}

For a set of points $\P$, we show next, that the locality graph
$\GL=\GL(\P, \eps/32, 1/8)$ is an $(1+\eps)$-spanner. Formally, for
two points $x, y \in \P$, such that $xy$ is an edge of $\GL$, its
weight is $\dY{x}{y}$. For any two points $x,y \in \P$, let
$\dGY{x}{y}$ denote the length of their shortest path in $\GL$. The
graph $\GL$ is \emphi{$(1+\eps)$-spanner}, if for all $x,y \in \P$, we
have $\dGY{x}{y} \leq (1+\eps)\dY{x}{y}$. In the following, for
$X, Y \subseteq \P$, let
$\dsGY{X}{Y} = \min_{x \in X, y \in Y} \dGY{x}{y}$.

\begin{theorem}
    \thmlab{eps_spanners} %
    Let $\P \subseteq \UC$ be a set of $n$ points, and let
    $\eps \in (0,1)$ be a parameter. The \defrefY{locality
       graph}{locality:graph} $\GL = \GL(\P,\eps/32,1/8)$ is an
    $(1+\eps)$-spanner of $\P$. The graph $\GL$ has
    $O( n \Eps^{d-1} \log \Eps)$ edges. Updating $\GL$, after an
    insertion or a deletion of a point, takes
    $O \bigl( (\Eps^{d-1}\log^2 \Eps) \log n \bigr)$ time. Each such
    operation causes at most $O(\Eps^{d-1} \log \Eps)$ edges to be
    removed from or inserted into $\GL$, and this also bounds the
    maximum degree of a vertex of $\GL$.
\end{theorem}
\begin{proof}
    Let $\Pi$ be the set of orderings used to construct $\GL$ --
    specifically, it is the one provided by \thmref{conclusion} so
    that the orderings of $\Pi$ are $(\eps/32, 1/8)$-local.

    We prove the spanner property by induction on the distance of
    pairs of points of $P$. It is easy to verify, arguing as in
    \thmref{bichromatic_closet_pair}, that the closest pair of points
    of $\P$ are connected by an edge in the locality graph $\GL$,
    which establish the base of the induction.

    Fix a pair $p, q \in \P$ and assume by the induction hypothesis
    that for all pairs $x, y \in \P$, such that
    $\dY{x}{y} < \dY{p}{q}$, we have that
    $\dGY{x}{y} \leq (1 + \eps)\dY{x}{y}$.  Let $\order \in \Pi$ be
    the order in $\Pi$ that is $(\eps/32, 1/8)$-local for $p$ and $q$,

    \begin{figure}[h!]
        \centerline{\includegraphics{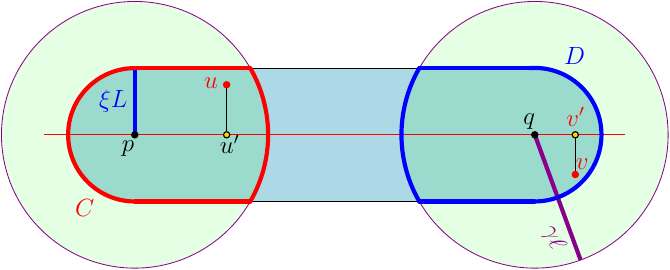}}
        \caption{Illustration for the proof of \thmref{eps_spanners}.}
        \figlab{e:s}
    \end{figure}

    Let $\numA = \dY{p}{q}$. The set $\P' = \P \cap \order[p,q]$ is
    contained in the union of the two sets
    \begin{equation*}
        C= \ballY{p}{ \gamma \numA } \cap \hippoY{pq}{ \epsA \numA \bigr. }
        \qquad\text{and}\qquad%
        D= \ballY{q}{ \gamma \numA} \cap \hippoY{p q}{ \epsA \numA \bigr.},
    \end{equation*}
    where $\gamma = 1/8$ and $\epsA = \eps/32$, see \figref{e:s} and
    \figref{locality} (iii).  We consider $\P'$ to be sorted by
    $\order$. There must be two consecutive points $u,v \in \P'$, such
    that $u \in C$ and $v \in D$. By construction, $uv$ is an edge of
    $\GL$. By the gap property, we have that
    \begin{equation}
        \dY{u}{v}
        \geq%
        \dsY{C}{D} = (1-2\gamma) \dY{p}{q} \geq (3/4)
        \numA.
        \eqlab{gap}
    \end{equation}

    Let $u', v'$ be the projections of $u, v$, respectively, on the
    line spanned by $p q$, see \figref{e:s}.
    We have the following:
    \begin{compactenumA}
        \item $\dY{u}{u'} \leq \epsA \numA$ and
        $\dY{v}{v'} \leq \epsA \numA$.

        \smallskip%
        \item
        $\dY{p}{u} \leq \dY{p}{u'} + \dY{u'}{u} \leq \dY{p}{u'} +
        \epsA \numA$.
        \smallskip
        \item $\dY{v}{q} \leq \dY{v'}{q} + \epsA \numA$.  \smallskip
        \item $\dY{u}{v} \leq \dY{u'}{v'} + 2\epsA \numA$.
        \smallskip%
        \item As $p, u', v', q$ are all on the same line (spanned by
        $p$ and $q$), and are contained in the hippodrome
        $\hippoY{\p\q}{\epsA \numA}$, the point $u'$ is closer to $p$
        than to $q$, and similarly $v'$ is closer to $q$ than to $p$.
        It follows hat
        $\dY{p}{u'} + \dY{u'}{v'} + \dY{v'}{q} \leq \dY{p}{q} + 4
        \epsA \numA$.
    \end{compactenumA}
    As such, by induction, we have
    \begin{align*}
      \dsGY{p}{q}
      &\leq
        \dsGY{p}{u} + \dsGY{u}{v} + \dsGY{v}{q}
        \leq%
        (1 + \eps)\dY{p}{u} + \dY{u}{v} + (1 + \eps)\dY{v}{q}
      \\%
      &=
        \underbrace{(1 + \eps)\bigl(\dY{p}{u} + \dY{u}{v} + \dY{v}{q}\bigr)}_{X} - \eps
        \dY{u}{v}.
    \end{align*}
    We have that
    \begin{align*}
      X &\leq
          (1 + \eps)(\dY{p}{u'} + \dY{u'}{v'} + \dY{v'}{q}  +4 \epsA
          \numA)
      \\%
        &\leq
          (1 + \eps)(\dY{p}{q} +8 \epsA \numA)
      \\&
      =
      (1 + \eps)(\numA  +8 \epsA \numA)
      \\&%
      \leq
      (1 + \eps)\numA + 16 \epsA \numA,
      \leq
      (1 + \eps)\numA +  \smash{\frac{1}{2}}\eps \numA,
    \end{align*}
    as $\epsA = \eps/32$.  The gap property, as stated in \Eqref{gap},
    implies that $\eps \dY{u}{v} \geq \tfrac{3}{4}\eps L$, and thus
    \begin{equation*}
      \dsGY{p}{q}
      \leq
      X -\eps \dY{u}{v}
      \leq%
      (1 + \eps)\numA + \frac{\eps}{2}  \numA  - \frac{3}{4}\eps \numA
      \leq
      (1 + \eps)\numA.
      \JoCG{\tag*{\qedhere}}
    \end{equation*}
\end{proof}

\begin{remark}
    Observe that the gap property is critical in making the proof of
    \thmref{eps_spanners} work -- we need a long ``bridge'' edge that
    the spanner uses, to charge the inductive error to.
\end{remark}

\section{Lower bounds on the number of orderings}
\seclab{lower:bounds}

\subsection{Lower bound of number of \LSOs with the
   \TPDF{$\eps$}{epsilon}-local property}

\begin{lemma}
    \lemlab{old:tight}%
    For any fixed integer $d$, and $\eps \in (0,1)$, there is a set
    $\P$ of $\Theta(\Eps^d)$ points in $\Re^d$, such that any set
    $\Pi$ of \LSOs of $\P$ with the $\eps$-local property, see
    \defref{local}, must be of size $\Omega(\Eps^d)$, where
    $\Eps = 1/\eps$.
\end{lemma}
\begin{proof}
    Let $m = \floor{\Eps / d }$.  Consider the integer grid
    $\P = \IRX{m}^d$.  Consider any pair of points $ p, q \in \P$, any
    other point $z \in \P$ is in distance at least one from $\p$ or
    $\q$. Setting, $\ell = \dY{\p}{q}$, and observing that
    $\eps \ell \leq m\sqrt{d} /d < 1$, we have that
    \begin{equation*}
        P \cap\pth{
           \ballY{\p}{\eps \ell}
           \cup
           \ballY{\q}{\eps \ell}
        }
        = \{ \p, \q\}.
    \end{equation*}
    Namely, there must $\exists \sigma \in \Pi$ such that $p$ and $q$
    are adjacent.  Thus, the number of orderings is at least
    \begin{equation*}
        \frac{\binom{\cardin{\P}}{2}}{\cardin{\P} -1}
        =%
        \frac{m^d}{2}
        =
        \Omega(\Eps^d).%
        \JoCG{\qedhere}%
    \end{equation*}
\end{proof}%

\subsection{Lower bound for \LSOs with the
   \TPDF{$(\eps,\gamma)$}{(epsilon,gamma)}-local property}

\begin{theorem}
    For any fixed integer $d$, and $\eps \in (0,1)$, there exists a
    set of point $P$, of size $\Omega(\Eps^{d-1})$, such that any set
    $\Pi$ of \LSOs for $\P$ with $(\eps,\gamma)$-local property must
    be of size $\Omega(\Eps^{d-1})$, for any $\gamma \geq \eps$.
\end{theorem}
\begin{proof}
    Let $\Sphere$ be the unit sphere centered at the origin $o$.  Let
    $N$ be a $4\eps$-packing of $\Sphere$. By \remref{packing:l:b},
    $\cardin{N} = \Theta( \Eps^{d-1})$. Let $P = N \cup \{ o
    \}$. Observe that for any point $p \in N$, there must be an order
    $\order \in \Pi$, such that $o$ and $\p$ are consecutive. Indeed,
    consider any order $\order \in \Pi$, and observe that the
    $\eps$-hippodrome of $o \p$ does not contain any other points of
    $\P$, which implies that they must be consecutive in one of the
    orderings. Since $|N| = \Omega(\Eps^{d-1})$, and every ordering
    can cover at most two pairs of points adjacent to $o$, it follows
    that $\cardin{\Pi} \geq \cardin{N}/2 = \Omega(\Eps^{d-1})$.
\end{proof}%

\subsection{Lower bound for \LSOs that their locality graph is a
   spanner}

Le and Solomon \cite{ls-toes-19,ls-toes-19-arxiv} showed that any
$\eps$-spanner in $\Re^d$ of $n$ points has $\Omega( n/ \eps^{d-1})$
edges. Each \LSO gives rise to $n-1$ edges in the locality graph. As
such, one needs at least $\Omega( 1 / \eps^{d-1})$ \LSOs to construct
an $\eps$-spanner using the locality graph.

\begin{lemma}
    Any set of \LSOs that their locality graph is an
    $(1+\eps)$-spanner has size $\Omega(1/\eps^{d-1})$.
\end{lemma}

\JoCG{%
   \paragraph*{Acknowledgements.}
}%
\NotJoCG{%
   \paragraph*{Acknowledgements.}
}%
The authors thank the anonymous referees for their detailed and
insightful comments.

\printbibliography

\end{document}